# Extending the Measurement of Composite Indicators Towards a Non-convex Approach: Corporate Social Responsibility for the Food and Beverage Manufacturing Industry

Magdalena Kapelko[a*] and Lidia Ortiz[b]

[a] Wroclaw University of Economics and Business, Department of Logistics, Poland

[b] Center of Operations Research (CIO). Miguel Hernandez University of Elche (UMH), Elche (Alicante), Spain

* Corresponding author: magdalena.kapelko@ue.wroc.pl (Tel.: +48713680479). Department of Logistics, Wroclaw University of Economics and Business, ul. Komandorska 118/120, 53-345, Wroclaw, Poland

*/December 2022/*

*Abstract:*

This paper computes composite indicators of corporate social responsibility (CSR) from an efficiency perspective for food and beverage manufacturing firms in various world regions over the period 2011–2018. From a methodological perspective, we extend the measurement of composite indicators within data envelopment analysis, allowing for the non-convexities of the production set and for the appropriate comparison of indicators between groups of firms. From an empirical point of view, we contribute by comparing the efficiency in CSR practices of food and beverage companies across regions of Europe, the United States and Canada, and Asia-Pacific. The study reveals differences in CSR efficiency between food and beverage firms in the regions considered, with USA-Canadian firms tending to perform best, followed by European firms, and Asian-Pacific firms achieving the worst efficiency results. The study also shows that regional catching up in performance occurred over the analyzed period.

# 1. Introduction

Corporate social responsibility (CSR) can be defined as an integration of social, environmental, ethical, human rights, and consumer concerns into firms' business operations and strategy (European Commission, 2011). Its essence lies in the fostering of social good beyond economic interests of the firm and beyond legal requirements (McWilliams and Siegel, 2001). Six key terms underpin the CSR concept: economic, social, ethical, stakeholders, sustainability, and voluntary (Sarkar and Searcy, 2016). In the food and beverage manufacturing industry, CSR engagement is increasing in importance because of this sector's strong impact on the economy, the environment, and the society (Hartmann, 2011). In particular, the food and beverage industry faces CSR challenges related to factors such as food safety scares, which have made consumers more concerned with food safety, consumer awareness of the links between food and health and of the responsible consumption (impact on climate change, animal welfare or social and economic inequality) (European Commission, 2016).

CSR encompasses a broad range of activities within its three main dimensions: social, environmental and governance CSR. Social CSR is a corporate response to the needs of primary stakeholders and includes labor relations (such as good workplace conditions or employee profit sharing), employment diversity (such as employment of minorities or disabled persons), business practice (such product safety or quality), and community involvement (such as charitable donations or education programs). Environmental CSR reflects firms' environmental awareness: involvement in recycling, waste reduction, renewable energies, or biotechnology. Governance dimension of CSR encompasses exceptional practices with board independence, auditor independence, or shareholders rights. One of the major challenges in the research on CSR is the aggregation of these CSR dimensions into overall measures of CSR practices by firms. This aggregation is undertaken in the literature using different methods such as summing up the scores on different CSR dimensions, subjectively weighting different dimensions, or introducing objective weighting through data envelopment analysis (DEA). DEA (Banker et al., 1984; Charnes et al., 1978) is a methodology based on mathematical programming for the measurement of efficiency of decision making units (DMUs) that convert multiple inputs into multiple outputs. DEA constructs a piece-wise frontier enveloping the data and determines the distance of each DMU to this frontier. The DEA-based approach to aggregate indicators is called the benefit-of-the-doubt (BoD) DEA model (Cherchye et al., 2007a,b; Cherchye et al., 2004), which measures efficiency with inclusion of indicators as outputs and a unitary input with the aim of creating an overall and objective aggregated indicator for each analyzed unit (a so-called composite indicator). It presents a certain advantage in that the weights to aggregate CSR metrics are derived directly from the data. Because of this characteristic, this study builds upon BoD model to derive composite indicators of CSR.

The use of BoD model in CSR research was initiated by Bendheim et al. (1998), who assessed the efficiency of CSR with respect to five key stakeholders using data concerning light manufacturing, consumer products, primary industry, service, heavy industry and transportation in the USA. Chen and Delmas (2011) computed CSR indicators for US companies in manufacturing, finance, insurance and real estate, services, retail trade, mining, transportation, wholesale trade, construction, public administration, and agriculture, forestry and fishing, while Belu and Manescu (2013) analyzed CSR of nonfinancial large publicly traded companies across sectors of basic resources, industrial goods, industrial services, consumer search goods, durable experience goods, nondurable experience goods, and experience services. Staessens et al. (2019) used BoD for CSR of Flemish sheltered workshops.

Aparicio and Kapelko (2019) extended the BoD model and applied it to the CSR data of US companies in construction, finance, manufacturing, mining, retail trade, services, transportation, and wholesale trade sectors. Oliviera et al. (2019) proposed an innovative BoD composite indicator and applied it to the CSR of large mining firms. Lahouel et al. (2021 and 2022) provided further extensions of the BoD model and applied them to international airline companies. Within studies that have focused exclusively on the food and beverage manufacturing, the BoD model was applied in Engida et al. (2018) and extended in Aparicio et al. (2020) to analyze CSR of these firms in Europe.[1] All of these studies, pertaining to different industries and specifically referring to food and beverage manufacturing, are restricted to firms in specific geographical regions and, as far as we are aware, no previous contribution has computed CRS composite indicators for food and beverage manufacturing industry in different geographical regions worldwide.

From a methodological perspective, many different extensions of the BoD model have been developed. Examples include the non-radial model with slacks (Sahoo and Acharya, 2012); the directional model, which takes into account the preference structure among indicators (Fusco, 2015); the directional model, which accounts for undesirable indicators (Zanella et al., 2015); the robust and non-compensatory composite indicator within the directional model (Vidoli et al., 2015); the robust BoD model, which considers external factors (Fusco et al., 2018); an intra- and inter-group BoD models (Karagiannis and Karagiannis,

---

[1] Another strand of literature uses DEA models in the standard production framework, in which inputs are converted into outputs, to assess the overall firms' performance with the inclusion of CSR factors. For example, Belu (2009) studied large corporations listed on the world's main stock exchanges; Lee and Saen (2012) looked at the Korean electronics industry; Puggioni and Stefanou (2019) examined worldwide food and beverage manufacturing firms; Chambers and Serra (2018) studied a sample of global firms; Engida et al. (2020) investigated European food and beverage manufacturing firms; Ait Sidhoum et al. (2020) looked at Catalan arable crop farms, Kapelko et al. (2021) examined European firms in capital, consumption and other industries; and Kapelko and Oude Lansink (2022) studied US food and beverage manufacturing companies.

2018); a translation-invariant directional distance function model (Aparicio and Kapelko, 2019); a model based on directional distance function and goal programming (Oliviera et al., 2019); a model assuming a least distance to the frontier (Aparicio et al., 2020); a method integrating the spatial dependence into the robust directional model in the case of undesirable outputs (Fusco et al., 2020); a model based on the notion of multiple attribute utility theory (Lahouel et al. 2021); a model based on directional distance function and the Malmquist index (Lahouel et al. 2022); and a model that combines BoD with multi-directional efficiency analysis (Fusco, 2022)[2]. All of the BoD developments assumed convexity of the production set, which means that points on the frontier used to evaluate observations are constructed based on linear combinations of actual observations, and not on actual observations themselves.[3] However, such an assumption is often unrealistic since some of such observations can never actually be realized, and recent evidence shows that production set is often non-convex (e.g., Kerstens et al., 2019; Wilson, 2021). The direct empirical implication is non-convexity and the non-convexifying free disposal hull (FDH) estimator introduced by Deprins et al. (1984) as the approach to data envelopment. However, to the best of our knowledge, the FDH approach has never been considered in the context of composite indicators created by BoD. Moreover, it is often important to assess differences in efficiencies between groups of firms that could be associated, for example, with regional location of firms, as in the case of our empirical analysis. It is well known that efficiency measures cannot be directly compared between these groups due to the existence of different technologies; specific approaches should be used in such contexts, such as the so-called Camanho-Dyson index (Camanho and Dyson, 2006). Camanho and Dyson (2006) introduced a Malmquist-type index that derives a relative performance of two or more groups of DMUs, allowing for direct efficiency comparisons of groups of DMUs. Later, Aparicio et al. (2017) extended the original Camanho-Dyson index to the context of the panel data or the pseudo-panel data with the objective of assessing the changes in the performance gap over time. Aparicio and Santin (2018) introduced the improvement on Aparicio et al. (2017) by providing the solutions to interpretational and non-circularity problems. The original paper of Camanho and Dyson (2006), as well as further developments of the Camanho-Dyson index in the technical efficiency context (Aparicio et al., 2017; Aparicio and Santin, 2018; Aparicio et al., 2021[4]), assumed constant returns to scale (CRS) or variable returns to scale (VRS). However, the Camanho-Dyson index has never been used in the context of the FDH approach.

---

[2] The applications of BoD model, in addition to CSR, include, for example, human development (Van Puyenbroeck and Rogge, 2020), competitiveness (Lafuente et al., 2022), waste management (Fusco et al., 2018), hospitals (Matos et al., 2021), international economic diplomacy (Charles et al., 2022), and quality of life (Rogge and Van Nijverseel, 2019).
[3] Specifically, convexity of a production set implies that if two input vectors x1 and x2 can produce two output vectors y1 and y2, respectively, then any positive linear combination of these input vectors $\alpha x1 + (1-\alpha)x2$ with $\alpha \in [0, 1]$, can produce the output vector $\alpha y1 + (1-\alpha)y2$ (Kerstens et al., 2019).

The present study aims to fill in the gaps in the literature outlined above. We constructed composite indicators in the context of CSR from an efficiency perspective using a BoD model with non-convex structure within the Camanho-Dyson index, extended with recent developments by Aparicio et al. (2017) and Aparicio and Santin (2018). The present study is the first to formulate a BoD model with non-convexity assumption. The empirical focus is a sample of food and beverage manufacturing firms in different world regions over the period 2009–2018. The study contributes to the literature in two ways. From an empirical point of view, this paper is the first to analyze and compare efficiency in CSR practices by food and beverage companies across regions of Europe, the United States and Canada, and the Asia-Pacific. Second, methodologically, we extend the measurement of composite indicators using

---

[4] The extension of the Camano-Dyson index towards cost efficiency measurement was put forward by Thanassoulis et al. (2015).

DEA, allowing for the non-convexities of the production set and for the appropriate comparison of these indicators between groups of firms.

The remainder of this paper is organized as follows. The next section describes the methodology to compute composite indicators and to compare these indicators between groups of DMUs. The subsequent section provides information on the dataset, followed by the description and interpretation of the results. The final section concludes.

**2. Methods**

**2.1. BoD Model Assuming FDH**

We used a BoD model to aggregate CSR scores across three dimensions of environmental, social and governance CSR, and measure efficiency in CSR (Cherchye et al., 2007a,b; Cherchye et al., 2004). We extended this model to assume non-convexities through the FDH approach. Stochastic Frontier Analysis (SFA) (Aigner et al., 1977) could not be used in this context since it is not able to account for non-convexities. Also, composite indicators cannot be derived using SFA.[5]

The original BoD model is a DEA model without inputs, where all of the subindicators are treated as outputs to generate an overall and objective aggregated indicator for each assessed firm. The usage of DEA provides an advantage in that it avoids any possible controversies related to the selection of weights for each specific indicators, since weighting has objective nature. The linear programming model to construct composite indicator for DMU 0 is given below (Cherchye et al., 2004):

---

[5] Alternatively, one could consider the usage of Stochastic Non-Smooth Envelopment of Data (StoNED) method (Kuosmanen and Kortelainen, 2012). However, this model has never been extended to the context of composite indicators, so its usage in our context would require further investigations.

$$CI_0 = \text{Max} \sum_{r=1}^{s} w_r y_{r0}$$

s.t.

$$\sum_{r=1}^{s} w_r y_{rj} \leq 1, \quad j=1,...,n, \tag{1}$$

$$w_r \geq 0, \quad r=1,...,s$$

In Model (1), $y_{rj}$ corresponds to the value of the output indicator $r$ ($r=1,...,s$) in DMU $j$ ($j=1,...,n$), $y_{r0}$ indicates the value of the output indicator $r$ ($r=1,...,s$) in DMU $0$ under evaluation, while $w_r$ ($r=1,...,s$) signifies weights. The essence of this model is to find weights $w_r$ that maximize the composite indicator for DMU $j$. The constraint of the model forces the value of the composite indicator to be less or equal to 1. If DMU $j$ is on the frontier, then the objective value will be equal to 1, which means that no other DMU will be able to obtain a higher weighted sum with the most favorable set of weights for DMU $j$. Otherwise, for other DMUs the objective function will be lower than 1, which means that, even with the most favorable set of weights there are other DMUs that obtain a higher weighted sum.

Model (1) is equivalent to an input-oriented CRS DEA model in multiplier form, with all subindicators treated as outputs and a single input equal to one (Van Puyenbroeck, 2018). As Van Puyenbroeck (2018) showed, it is equivalent to a reciprocal of the output-oriented VRS model without the inputs of Lovell at al. (1995) and Lovell and Pastor (1999, 1997). Therefore, Model (1) assumes convexity. In order to relax convexity, we formulated Model (1) assuming FDH. Based on the findings of Leleu (2006) that formulated primal and dual models for FDH, we can define the BoD model with a non-convex structure as follows:

$$CI_{NC,0} = \underset{w_{rj},\pi_0}{Max} \quad \pi_0$$

$$s.t.$$

$$\sum_{r=1}^{s} w_{rj}\left(y_{rj} - y_{r0}\right) + \pi_0 - 1 \leq 0, \quad j = 1,...,n$$
$$\sum_{r=1}^{s} w_{rj} y_{rj} - 1 = 0, \quad j = 1,...,n, \quad (2)$$
$$w_{rj} \geq 0, \quad r = 1,...,s,$$

Alternatively, a FHD BoD model can be derived based on the premise that the BoD model is a reciprocal of the output-oriented VRS model without inputs, for which the formulation of the FDH version is straightforward.

Figure 1 illustrates a BoD model that both assumes convexity and does not assume convexity (FDH BoD model) for the example of two indicators Y1 and Y2, which could be represented by the CSR variables, as in our empirical application. The implication of the convexity assumption can be observed with Observation B, which is efficient relative to the FDH frontier, but is inefficient relative to the convex combination of A and C on the DEA model. Assuming the FDH approach, the inefficient Observation D is projected onto Point D' situated on the orthant spanned by B, which is one of the dominating observations. Unit B functions as a role model for the inefficient Unit D. In DEA, inefficient observations are projected onto a fictitious linear combination of efficient observations. For example, Observation D is projected to Point D", which is a linear combination of Observations A and C, which in DEA functions as a role model. Hence, FDH identifies one role model, which is an actual observation in contrast to a fictitious convex combination of efficient producers in DEA. In the FDH approach, the efficiency measure for Unit D is defined by 0D/0D', while in DEA it is measured by 0D/OD''. Moreover, in FDH, Observations A, B, and C are efficient because they are not dominated by any other observation. D is inefficient because it produces fewer of both outputs than B, so it is dominated by B. Observations A and C do not dominate any other observations – they are efficient by default. Observation B is efficient and not dominated in FDH, but imposing convexity it is dominated by some convex combination of A and C.

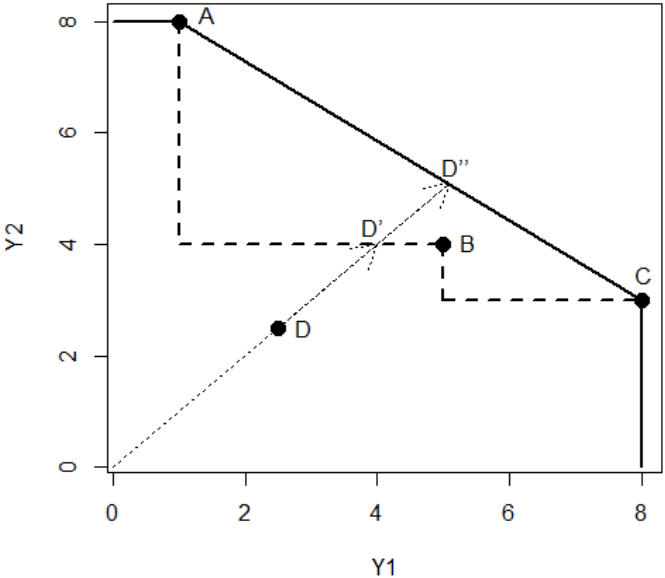

**Fig. 1.** The illustration of the BoD FDH model.

**2.2. Camanho-Dyson index using the BoD approach assuming FDH**

Let us assume that we have observed $n^{A^p}$ DMUs in Group A in Period $p$, $p = t, t+1$, with the value of the output indicator $Y^{A^p} \in R_+^s$ and that we have also observed $n^{B^s}$ DMUs in Group B in Period $p$, $p = t, t+1$, with the value of the output indicator $Y^{B^p} \in R_+^s$.

We denoted $D^{R^h}\left(Y_k^{A^p}\right)$ as the Shephard input distance function (Shephard, 1953) calculated from observation $\left(Y_k^{A^p}\right)$ in Group A in Period $p$, to the frontier of the technology of Reference Group R in period $h$, $T_C^{R^h}$. A similar notation is used for the distance for a unit in B with respect to the technology of Reference Group R, and for the distance from a unit that belongs to the same group as the reference technology. From Model (2) we can define the composite indicators for Group A (3) and for Group B (4) in period $p$, to the technology frontier of the reference Group R in period $h$.

$$1/D^{R^h}(Y_0^{A^p}) = CI_{NC,0}^{A^p} = \underset{w_{rj}^{A^s},\pi_0^s}{Max} \quad \pi_0^p$$

$$s.t.$$

$$\sum_{r=1}^{s} w_{rj}^{A^p}\left(y_{rj}^{R^h} - y_{r0}^{A^p}\right) + \pi_0^p - 1 \le 0, \quad j = 1,...,n^{A^p}$$

$$\sum_{r=1}^{s} w_{rj}^{A^p} y_{rj}^{R^h} - 1 = 0, \quad j = 1,...,n^{A^p},$$

$$w_{rj}^{A^p} \ge 0, \quad r = 1,...,s,$$

(3)

$$1/D^{R^h}(Y_0^{B^p}) = CI_{NC,0}^{B^p} = \underset{w_{rj}^{B^s},\pi_0^s}{Max} \quad \pi_0^p$$

$$s.t.$$

$$\sum_{r=1}^{s} w_{rj}^{B^p}\left(y_{rj}^{R^h} - y_{r0}^{B^p}\right) + \pi_0^p - 1 \le 0, \quad j = 1,...,n^{B^p}$$

$$\sum_{r=1}^{s} w_{rj}^{B^p} y_{rj}^{R^h} - 1 = 0, \quad j = 1,...,n^{B^p},$$

$$w_{rj}^{B^p} \ge 0, \quad r = 1,...,s,$$

(4)

Following Camanho and Dyson (2006) and Aparicio and Santín (2018) for comparisons between groups, we can then define the CD index with reference group ($R$) for the BoD model with non-convex structure, in order to assess the relative performance between A and B over a period of time:

$$CD_p^{AB}(R^h) = \frac{\left(\prod_{j=1}^{n^{B^p}} D^{R^h}\left(Y_j^{B^p}\right)\right)^{1/n^{B^p}}}{\left(\prod_{k=1}^{n^{A^p}} D^{R^h}\left(Y_k^{A^p}\right)\right)^{1/n^{A^p}}}. \tag{5}$$

In terms of the interpretation of $CD_p^{AB}(R^h)$, a value greater than one of the indicator indicates that Group A performed relatively better on average than Group B, and a value less than one indicates a better average performance of Group B compared to Group A.

$CD_p^{AB}(R^h)$ can be decomposed into the terms of efficiency spread ($EG_p^{AB}$) and technological gap ($TG_p^{AB}(R^h)$):

$$CD_p^{AB}(R^h) = \underbrace{\frac{\left(\prod_{j=1}^{n^{B^p}} D^{B^p}\left(Y_j^{B^p}\right)\right)^{1/n^{B^p}}}{\left(\prod_{k=1}^{n^{A^p}} D^{A^p}\left(Y_k^{A^p}\right)\right)^{1/n^{A^p}}}}_{EG_p^{AB}} \cdot \underbrace{\left[\frac{\left(\prod_{j=1}^{n^{B^p}} D^{R^h}\left(Y_j^{B^p}\right)\right)^{1/n^{B^p}}}{\left(\prod_{j=1}^{n^{B^p}} D^{B^p}\left(Y_j^{B^p}\right)\right)^{1/n^{B^p}}} \cdot \frac{\left(\prod_{k=1}^{n^{A^p}} D^{A^p}\left(Y_k^{A^p}\right)\right)^{1/n^{A^p}}}{\left(\prod_{k=1}^{n^{A^s}} D^{R^h}\left(Y_k^{A^s}\right)\right)^{1/n^{A^p}}}\right]}_{TG_p^{AB}(R^h)}. \tag{6}$$

The first ratio shows the technical efficiency change ($EG_p^{AB}$), which measures the efficiency gap between both groups in the Time Period $p$. The second measure ($TG_p^{AB}(R^h)$) evaluates the productivity gap between the frontiers of the two analyzed groups, A and B, in the Time Period $p$, measured on the base reference technology ($R$) in the base Time Period $h$.

Regarding the relative performance difference between two groups over two time periods, Aparicio and Santín (2018) proposed the following index:

$$PPMI_{t,t+1}^{AB}(R^h) = \frac{CD_{t+1}^{AB}(R^h)}{CD_t^{AB}(R^h)} \qquad (7)$$

Following Aparicio et al. (2017), we include the interpretation of the PPMI index based on the results of the $CD_p^{AB}(R^h)$ in $t$ and $t+1$.

Setting 1: $CD_t^{AB}(R^h)$, $CD_{t+1}^{AB}(R^h) < 1$. This means that, on average, Group B had a better relative performance than Group A in both the $t$ and $t + 1$ periods. Regarding the value of PPMI, there are two possibilities.

    (1a) $PPMI_{t,t+1}^{AB}(R^h) < 1$, which means that the relative performance gap was opened up by B over A.

    (1b) $PPMI_{t,t+1}^{AB}(R^h) > 1$, which means that Group A is catching up to Group B.

Setting 2: $CD_t^{AB}(R^h)$, $CD_{t+1}^{AB}(R^h) > 1$. This means that Group A had a better relative performance than Group B in both $t$ and $t + 1$. Regarding PPMI, we have again two scenarios.

    (2a) $PPMI_{t,t+1}^{AB}(R^h) < 1$, which means that Group B is catching up to Group A.

    (2b) $PPMI_{t,t+1}^{AB}(R^h) > 1$, which means that the relative performance gap was opened up by A over B. Setting 3: $CD_t^{AB}(R^h) > 1$ and $CD_{t+1}^{AB}(R^h) < 1$. Under this scenario, Group A had a better relative performance than Group B in Period $t$, but B had a better relative performance

than A in the second Period, *t + 1*. The value of PPMI is: $PPMI_{t,t+1}^{AB}(R^h) < 1$. In this case, Group A worsened drastically from Period *t* to Period *t + 1*.

Setting 4: $CD_t^{AB}(R^h) < 1$ and $CD_{t+1}^{AB}(R^h) > 1$. In this case, B had a better relative performance than A in Period *t*, but the relation between both groups changed drastically in Period *t + 1*, where A had a better relative performance than B. Regarding the PPMI, we have again only one possibility: $PPMI_{t,t+1}^{AB}(R^h) > 1$. The status of Group A improved from Period *t* to Period *t + 1*.

The $PPMI_{t,t+1}^{AB}(R^h)$ can also be decomposed into efficiency gap change (EGC) and technological gap change (TGC) in the following way:

$$PPMI_{t,t+1}^{AB}(R^h) = EGC_{t,t+1}^{AB} \cdot TGC_{t,t+1}^{AB}(R^h) \qquad (8)$$

Where:

$$EGC_{t,t+1}^{AB} = \frac{EG_{t+1}^{AB}}{EG_t^{AB}} \qquad (9)$$

$$TGC_{t,t+1}^{AB}(R^h) = \frac{TG_{t+1}^{AB}(R^h)}{TG_t^{AB}(R^h)} \qquad (10)$$

### 3. Dataset

Our study is based on the CSR data obtained from Sustainalytics, a global leader in environmental, social and governance research and ratings (see www.sustainalytics.com). Unlike most other CSR data providers, Sustainalytics does not restrict itself to a specific geographic region, which makes it suitable for the purpose of this study, which is to compare firms in different regions[6]. Sustainalytics' dataset has been widely used in recent research (for example, Engida et al., 2018; Auer, 2016; Kim et al., 2016). The information on CSR indicators in Sustainalytics comes from multiple sources such as annual reports, CSR reports, CSR websites, press releases, local newspapers, or relevant websites (Auer, 2016). Firms in the food and beverage manufacturing industry were distinguished using the industry classification of Sustainalytics, which is based on Global Industry Classification Standard (GICS), in which this industry belongs to the wider sector of consumer staples (in which remaining industry groups are food and staples retailing, and household and personal products).

---

[6] Another source of CSR data that is not restricted to a specific region is Thomson Reuters. This data were recently applied in the study by Rajesh et al. (2022).

The Sustainalytics dataset includes scores on different indicators that capture three dimensions of CSR: environmental (for example, involvement in recycling, waste reduction and renewable energies), social (for example, employee profit sharing, product safety, employment of minorities and charitable donations), and governance (for example, board independence and shareholder rights). The dataset provides the information on two types of scores: raw scores and weighted scores. The raw scores range from 0 to 100, where 0 is the lowest rating. The weighted scores reflect the importance of an indicator in a specific industry – we use these scores in our analysis. The data availability in Sustainalytics allows us to analyze the 2009–2018 period covering the food and beverage companies from Europe, the United States and Canada, and the Asia-Pacific region.[7] Such regional classification is consistent with Sustainalytics and previous research (e.g., Auer, 2016). Due to the data limitation of Sustainalytics, it was impossible to undertake analysis by country. Nevertheless, the analysis by region is consistent with previous research that applied the same dataset (e.g., Engida et al., 2018; Auer, 2016).

---

[7] Europe includes Austria, Belgium, Cyprus, Denmark, France, Germany, Greece, Ireland, Italy, Luxembourg, Netherlands, Norway, Slovenia, Spain, Sweden, Switzerland, Ukraine and the UK. Asia-Pacific contains Australia, China, Hong Kong, India, Indonesia, Japan, Malaysia, New Zealand, Papua New Guinea, Philippines, Singapore, South Korea, Sri Lanka, Taiwan, Thailand, and Vietnam.

Table 1 reports basic descriptive statistics of CSR scores (averages and standard deviations) for our sample sub-divided by geographic regions. The data in the table indicate that, on average, the largest scores on social and environmental dimensions of CSR are achieved by European food and beverage firms, while firms in the United States and Canada score the largest values on governance dimension. Standard deviations relatively to their respective averages for all variables are relatively low.

*Table 1.* *Arithmetic means and standard deviations of CSR variables per region and overall, 2009–2018.*

| Region | Social | Environmental | Governance | No of observations |
|---|---|---|---|---|
| Asia-Pacific | 50.8193 (7.4162) | 47.9263 (12.0016) | 58.3950 (8.4447) | 745 |
| Europe | 59.4356 (10.979) | 59.9524 (12.6184) | 65.5308 (9.7374) | 464 |
| USA-Canada | 55.0590 (10.4169) | 58.9564 (12.4522) | 67.7492 (8.1960) | 418 |
| Overall | 54.3658 (10.027) | 54.1898 (13.576) | 62.8333 (9.7042) | 1627 |

## 4. Results and Discussion

First of all, we tested whether our sample is suitable for the analysis with FDH approach by applying the test of convexity. Kneip et al. (2016) developed a statistical test to assess the convexity in DEA based on the single sample split. Simar and Wilson (2020) provided further improvement in this test in the form of multiple sample splits, which remove ambiguity surrounding the choice of a single split. In both cases, the null hypothesis of convexity (that is, that both FDH and DEA estimators are consistent) is tested against an alternative one (that is, only FDH estimator is consistent). In the paper, we relied on the test by Simar and Wilson (2020) in order to check whether, for our data, only the proposed FDH estimator is consistent. The test was applied with 10 sample splits and 1000 bootstrap replications. Table 2 summarizes the results of convexity test for CSR indicator per year.[8]

*Table 2. Results of convexity test for CSR indicator.*

|  | Statistic | p-value |
| --- | --- | --- |
| 2009 | 0.5161 | 0.0690 |
| 2010 | 0.1579 | 0.6010 |
| 2011 | 0.3881 | 0.4940 |
| 2012 | 0.2017 | 0.5770 |
| 2013 | 0.8408 | 0.0030 |
| 2014 | 0.5535 | 0.0120 |
| 2015 | 0.5714 | 0.0040 |
| 2016 | 0.8812 | 0.0000 |
| 2017 | 0.5582 | 0.0170 |
| 2018 | 0.3455 | 0.5520 |

Table 2 reveals that the test does not reject convexity in 2010, 2011, 2012 and 2018, but it does reject convexity in 2009, 2013, 2014, 2015, 2016 and 2017. Hence, results provide a substantial evidence of non-convexity. Moreover, not rejecting null hypothesis does not mean that the null hypothesis is true. Also, the FDH estimator is consistent (Wilson, 2021). Therefore, our dataset seems to provide evidence of substantial non-convexity, and it is therefore suitable for the estimations of non-convex CSR indicator to reach the purpose of this paper.

---

[8] The results of the tests per region can be obtained upon request.

Next we computed a Camanho-Dyson index (Camanho and Dyson, 2006), enhanced with recent proposals by Aparicio et al. (2017) and Aparicio and Santin (2018), based on the developed CSR composite indicator with an FDH assumption. As explained, the application of this index requires a choice of the baseline technology (that is base-group and base-period) for the comparison of three regions over time. We were not able to use one of the regions from the sample for that purpose, since no clear evidence exists in the literature that any of the regions considered is best performing with regard to CSR. Instead we decided to choose the reference region-year using simple random sampling.[9]

Table 3 presents the baseline technology Camanho-Dyson index (5) and its components of efficiency spread gap and technological gap (6) for each year of the 2009–2018 period, as well as the average over the 2009–2018 period for food and beverage manufacturing firms in each of the regions considered (Europe, the United States and Canada, and the Asia-Pacific).[10]

---

[9] Simple random sampling considered a population of all units from all years; that is, 1627 observations in total. It assumed the precision of 5 percent, asymptotic normal confidence interval with the correction for finite populations two-sided at 95 percent, p=q=0.5. It proved that the necessary sample size to include is 311 units. Hence, the reference group was 311 units drawn randomly without a replacement, considering all groups and years.

[10] The application of Camanho-Dyson index can result in infeasible observations. Nevertheless, we did not encounter any infeasibilities in our empirical analysis.

*Table 3. Camanho-Dyson CSR index, and its decomposition into efficiency spread and technological gap, across years and regions*

|  | Camanho-Dyson index | | | Efficiency spread | | | Technological gap | | |
|---|---|---|---|---|---|---|---|---|---|
| Year and Region | A | B | C | A | B | C | A | B | C |
| **2009** | | | | | | | | | |
| A | 1.0000 | 1.0054 | 1.0207 | 1.0000 | 1.0301 | 0.9918 | 1.0000 | 0.9760 | 1.0291 |
| B |  | 1.0000 | 1.0153 |  | 1.0000 | 0.9629 |  | 1.0000 | 1.0544 |
| C |  |  | 1.0000 |  |  | 1.0000 |  |  | 1.0000 |
| **2010** | | | | | | | | | |
| A | 1.0000 | 0.9280 | 0.9125 | 1.0000 | 1.0348 | 0.9949 | 1.0000 | 0.8968 | 0.9172 |
| B |  | 1.0000 | 0.9833 |  | 1.0000 | 0.9615 |  | 1.0000 | 1.0227 |
| C |  |  | 1.0000 |  |  | 1.0000 |  |  | 1.0000 |
| **2011** | | | | | | | | | |
| A | 1.0000 | 0.9008 | 0.8788 | 1.0000 | 0.9537 | 0.8890 | 1.0000 | 0.9445 | 0.9885 |
| B |  | 1.0000 | 0.9756 |  | 1.0000 | 0.9321 |  | 1.0000 | 1.0466 |
| C |  |  | 1.0000 |  |  | 1.0000 |  |  | 1.0000 |
| **2012** | | | | | | | | | |
| A | 1.0000 | 0.8792 | 0.8612 | 1.0000 | 0.9208 | 0.8858 | 1.0000 | 0.9548 | 0.9722 |
| B |  | 1.0000 | 0.9796 |  | 1.0000 | 0.9620 |  | 1.0000 | 1.0183 |
| C |  |  | 1.0000 |  |  | 1.0000 |  |  | 1.0000 |
| **2013** | | | | | | | | | |
| A | 1.0000 | 0.8781 | 0.8587 | 1.0000 | 0.9415 | 0.9537 | 1.0000 | 0.9327 | 0.9003 |
| B |  | 1.0000 | 0.9779 |  | 1.0000 | 1.0130 |  | 1.0000 | 0.9654 |
| C |  |  | 1.0000 |  |  | 1.0000 |  |  | 1.0000 |
| **2014** | | | | | | | | | |
| A | 1.0000 | 0.8587 | 0.8558 | 1.0000 | 0.9720 | 0.9402 | 1.0000 | 0.8835 | 0.9103 |
| B |  | 1.0000 | 0.9967 |  | 1.0000 | 0.9673 |  | 1.0000 | 1.0304 |
| C |  |  | 1.0000 |  |  | 1.0000 |  |  | 1.0000 |
| **2015** | | | | | | | | | |
| A | 1.0000 | 0.8813 | 0.8622 | 1.0000 | 0.9869 | 0.9137 | 1.0000 | 0.8930 | 0.9436 |
| B |  | 1.0000 | 0.9783 |  | 1.0000 | 0.9258 |  | 1.0000 | 1.0567 |
| C |  |  | 1.0000 |  |  | 1.0000 |  |  | 1.0000 |
| **2016** | | | | | | | | | |
| A | 1.0000 | 0.8882 | 0.8833 | 1.0000 | 1.0168 | 0.9439 | 1.0000 | 0.8735 | 0.9359 |
| B |  | 1.0000 | 0.9945 |  | 1.0000 | 0.9283 |  | 1.0000 | 1.0713 |
| C |  |  | 1.0000 |  |  | 1.0000 |  |  | 1.0000 |
| **2017** | | | | | | | | | |
| A | 1.0000 | 0.8618 | 0.8891 | 1.0000 | 1.0320 | 1.0647 | 1.0000 | 0.8351 | 0.9012 |
| B |  | 1.0000 | 1.0317 |  | 1.0000 | 0.9560 |  | 1.0000 | 1.0791 |
| C |  |  | 1.0000 |  |  | 1.0000 |  |  | 1.0000 |
| **2018** | | | | | | | | | |
| A | 1.0000 | 0.8447 | 0.8939 | 1.0000 | 0.9457 | 1.0085 | 1.0000 | 0.8931 | 0.8864 |
| B |  | 1.0000 | 1.0583 |  | 1.0000 | 1.0664 |  | 1.0000 | 0.9924 |
| C |  |  | 1.0000 |  |  | 1.0000 |  |  | 1.0000 |
| **2009-2018** | | | | | | | | | |
| A | 1.0000 | 0.8926 | 0.8916 | 1.0000 | 0.9834 | 0.9586 | 1.0000 | 0.9083 | 0.9385 |
| B |  | 1.0000 | 0.9991 |  | 1.0000 | 0.9675 |  | 1.0000 | 1.0337 |
| C |  |  | 1.0000 |  |  | 1.0000 |  |  | 1.0000 |

A – Asia-Pacific, B – Europe, C – USA-Canada

The results in Table 3 show that, on average, for 2009–2018 the relative performance gap between Asia-Pacific and Europe was of 10.74 percent, in which European food and beverage firms perform better with regard to CSR than Asian-Pacific. Similarly, USA and Canadian firms scored better with regard to CSR than Asian-Pacific firms with the relative performance gap of 10.84 percent. The comparison of firms in Europe and USA-Canada over the 2009–2018 period shows a very small relative performance gap of 0.09 percent, according to which USA and Canadian firms had an edge over European firms concerning their CSR efficiency. Hence, in overall terms, USA-Canadian firms performed best with regard to CSR, followed by European firms, and finally Asian-Pacific firms scored lowest on their CSR efficiency. The decomposition of the overall measure shows that the better CSR performance of Europe over Asia-Pacific and of USA-Canada over Asia was due to the components of technical efficiency spread gap and technological gap; the latter played the major role, with 9.17 percent of the gap between Asia-Pacific and Europe and 6.15 percent of the gap between Asia-Pacific and USA-Canada. The better overall CSR performance of USA-Canada compared to Europe was caused by the component of the efficiency spread gap that is less dispersion in the technical efficiency levels of the firms in USA-Canada compared to Europe, despite Europe having an edge over USA-Canada in terms of the technological gap, which is the dominance of the best practice frontier of European firms.

Looking at specific years, the overall performance gaps ranged from 0.33 percent to 15.53 percent, with European and USA-Canadian food and beverage firms performing better in terms of CSR than Asian-Pacific firms for all years except 2009, and USA-Canadian firms having a productivity advantage over European firms in all years except 2009, 2017, and 2018. The performance gap between Asia-Pacific and Europe increased considerably till 2014 to the level of 14.13 percent, then decreased in 2015 and 2016 (to 11.87 percent and 11.18 percent, respectively), before increasing again in 2017 and 2018 (to 13.82 percent and 15.53 percent, respectively). The gap between Asia-Pacific and USA-Canada increased until 2014, up to 14.42 percent, and then decreased for the remaining years to 10.61 percent, indicating the productivity advantage of USA-Canada over Asia-Pacific with regard to CSR. Comparing USA-Canada with Europe, we see that the gap was rather small and almost constant over time, ranging from 0.33 percent to 5.83 percent. The main source of the gap between Asia-Pacific and Europe was a technological gap that favored Europe over Asia-Pacific in all years. The productivity gap between Asia-Pacific and USA-Canada can be explained by both components of efficiency spread gap and technological gap depending on the year. In particular, efficiency spread was the main source of the advantageous productivity gap of USA-Canada over Asia-Pacific in 2011, 2012, and 2015, while the technological gap was responsible for the advantage of USA-Canada over Asia-Pacific in the remaining years, except for 2009, in which the technological gap caused Asia-Pacific to have an edge over USA-Canada. Finally, the explanation of the productivity gap between USA-Canada and Europe was an efficiency spread gap in most of the years.

Table 4 presents the productivity gaps change over time (7), together with decomposition components of the efficiency spread gap change (9) and the technological gap change (10), for each pair of years and region, and average over the whole period.

*Table 4. Camanho-Dyson CSR index change, and its decomposition into efficiency spread change and technological gap change, across pairs of years and regions*

|  | Camanho-Dyson index change | | | Efficiency spread change | | | Technological gap change | | |
|---|---|---|---|---|---|---|---|---|---|
| Years and Region | A | B | C | A | B | C | A | B | C |
| 2009, 2010 | | | | | | | | | |
| A | 1.0000 | 0.9231 | 0.8940 | 1.0000 | 1.0046 | 1.0031 | 1.0000 | 0.9189 | 0.8912 |
| B |  | 1.0000 | 0.9685 |  | 1.0000 | 0.9985 |  | 1.0000 | 0.9699 |
| C |  |  | 1.0000 |  |  | 1.0000 |  |  | 1.0000 |
| 2010, 2011 | | | | | | | | | |
| A | 1.0000 | 0.9706 | 0.9630 | 1.0000 | 0.9217 | 0.8935 | 1.0000 | 1.0531 | 1.0778 |
| B |  | 1.0000 | 0.9921 |  | 1.0000 | 0.9695 |  | 1.0000 | 1.0234 |
| C |  |  | 1.0000 |  |  | 1.0000 |  |  | 1.0000 |
| 2011, 2012 | | | | | | | | | |
| A | 1.0000 | 0.9760 | 0.9801 | 1.0000 | 0.9654 | 0.9965 | 1.0000 | 1.0109 | 0.9835 |
| B |  | 1.0000 | 1.0042 |  | 1.0000 | 1.0321 |  | 1.0000 | 0.9729 |
| C |  |  | 1.0000 |  |  | 1.0000 |  |  | 1.0000 |
| 2012, 2013 | | | | | | | | | |
| A | 1.0000 | 0.9988 | 0.9970 | 1.0000 | 1.0225 | 1.0766 | 1.0000 | 0.9768 | 0.9261 |
| B |  | 1.0000 | 0.9982 |  | 1.0000 | 1.0529 |  | 1.0000 | 0.9480 |
| C |  |  | 1.0000 |  |  | 1.0000 |  |  | 1.0000 |
| 2013, 2014 | | | | | | | | | |
| A | 1.0000 | 0.9779 | 0.9967 | 1.0000 | 1.0324 | 0.9858 | 1.0000 | 0.9473 | 1.0110 |
| B |  | 1.0000 | 1.0192 |  | 1.0000 | 0.9549 |  | 1.0000 | 1.0673 |
| C |  |  | 1.0000 |  |  | 1.0000 |  |  | 1.0000 |
| 2014, 2015 | | | | | | | | | |
| A | 1.0000 | 1.0263 | 1.0074 | 1.0000 | 1.0154 | 0.9718 | 1.0000 | 1.0108 | 1.0366 |
| B |  | 1.0000 | 0.9816 |  | 1.0000 | 0.9571 |  | 1.0000 | 1.0256 |
| C |  |  | 1.0000 |  |  | 1.0000 |  |  | 1.0000 |
| 2015, 2016 | | | | | | | | | |
| A | 1.0000 | 1.0078 | 1.0245 | 1.0000 | 1.0302 | 1.0330 | 1.0000 | 0.9782 | 0.9918 |
| B |  | 1.0000 | 1.0166 |  | 1.0000 | 1.0027 |  | 1.0000 | 1.0139 |
| C |  |  | 1.0000 |  |  | 1.0000 |  |  | 1.0000 |
| 2016, 2017 | | | | | | | | | |
| A | 1.0000 | 0.9703 | 1.0066 | 1.0000 | 1.0150 | 1.0453 | 1.0000 | 0.9560 | 0.9629 |
| B |  | 1.0000 | 1.0374 |  | 1.0000 | 1.0299 |  | 1.0000 | 1.0073 |
| C |  |  | 1.0000 |  |  | 1.0000 |  |  | 1.0000 |
| 2017, 2018 | | | | | | | | | |
| A | 1.0000 | 0.9801 | 1.0054 | 1.0000 | 0.9164 | 1.0222 | 1.0000 | 1.0695 | 0.9836 |
| B |  | 1.0000 | 1.0258 |  | 1.0000 | 1.1154 |  | 1.0000 | 0.9196 |
| C |  |  | 1.0000 |  |  | 1.0000 |  |  | 1.0000 |
| 2009-2018 | | | | | | | | | |
| A | 1.0000 | 0.9812 | 0.9861 | 1.0000 | 0.9915 | 1.0031 | 1.0000 | 0.9767 | 0.9645 |
| B |  | 1.0000 | 1.0048 |  | 1.0000 | 1.0126 |  | 1.0000 | 0.9882 |
| C |  |  | 1.0000 |  |  | 1.0000 |  |  | 1.0000 |

A – Asia-Pacific, B – Europe, C – USA-Canada

The results confirm that, on average, for the whole period 2009–2018, European and USA-Canadian food and beverage companies had better relative performance than their Asian-Pacific counterparts. Moreover, on average for the whole period the relative performance gap was opened up by both Europe and USA-Canada over Asia-Pacific by 1.88 percent and 1.39 percent, respectively. Nevertheless, some improvement was taking place over the whole period since the productivity gap change was larger in value than the productivity gap itself for Asia-Pacific compared to Europe and USA-Canada. Looking at specific changes between periods we can observe that the realizations of Setting 1b, with Asia-Pacific catching up on both Europe and USA-Canada, were visible between 2014 and 2015, and 2015 and 2016, while between 2016 and 2017, and 2017 and 2018, this was the case for USA-Canada only. Setting 1a, in which the relative performance gap was opened up by both Europe and USA-Canada over Asia-Pacific, can be observed for the periods between 2010 and 2011, 2011 and 2012, 2012 and 2013, and 2013 and 2014, while Europe has opened the gap over Asia-Pacific also between 2016 and 2017, and 2017 and 2018. Setting 3, in which Asia worsened drastically compared to Europe and USA-Canada, occurred from period 2009 to 2010. Furthermore, although USA-Canada had a better relative performance than Europe, on average, Table 4 shows that Europe was clearly catching up for 2009-2018 by 0.48 percent. The periods for which a relative performance for Europe and USA-Canada was developing in favor of Europe were between 2011 and 2012, 2013 and 2014, 2014 and 2015, 2015 and 2016 (Setting 1b with catching up), between 2016 and 2017 (Setting 4), and 2017 and 2018 (Setting 2b in which the relative performance gap was opened up in favor of Europe). Between 2010 and 2011, and 2012 and 2013, a relative performance gap was opened up by USA-Canada over Europe (Setting 1a), while between 2009 and 2010 Europe worsened dramatically in comparison to USA-Canada (Setting 3).

On average, for 2009–2018, the productivity gap change between Asia-Pacific and Europe can be explained by both the efficiency spread gap change and technological gap change; nevertheless, these sources vary between specific periods. The gap change between Asia-Pacific and USA-Canada can be mostly explained by the technological gap change component overall for 2009–2018 as well as for all periods except 2010/2011, 2013/2014 and 2014/2015. Finally, the 2009–2018 the gap change between Europe and USA-Canada is due to efficiency spread gap change, which presents the same behavior for 2011/2012, 2012/2013, 2015/2016, 2016/2017, and 2017/2018.

To summarize, the results indicate that USA-Canadian food and beverage firms tended to achieve the best efficiency results with regard to CSR, followed by European firms, while Asian-Pacific firms proved to be the least CSR-efficient. The following explanations can be proposed for these results. Maigan and Ralston (2002) concluded that US firms were more willing to address CSR principles, processes, and issues than their European (French and Dutch) counterparts. In addition, USA has a much longer tradition of ethical components of CSR and socially responsible investing (Von Arx and Ziegler, 2014). These could serve as arguments for larger values of CSR efficiency for USA-Canadian firms than the European firms found in this study. Previous research findings confirm that CSR in Asia lags behind the best practice in European countries such as the UK (Chambers et al., 2003). Moreover, there is more activity in Europe and North America (excluding Mexico) than in Asia with regard to written internal CSR policies (Welford, 2005). For example, some Asian countries do not recognize the right of workers to standardized working hours, and it is a respected practice to remain at work for long hours and not be the first person the leave the office (Welford, 2005). This could help explain why Asia-Pacific lags behind Europe and USA-Canada in terms of CSR efficiency, as reported in the present study. Nevertheless, our findings also show that in some years of the analysis Asia-Pacific was catching up on Europe and USA-Canada, while Europe was catching up on USA-Canada. In fact, Welford (2005) demonstrated a new wave of interest in CSR in Asia, particularly among Japanese companies that are taking CSR seriously and increasing their CSR engagement.

To be able to assess the differences in CSR indices between regions with statistical precision, we ran a Simar and Zelenyuk (2006) test. We followed Vaz and Camanho (2012) to focus on indices consisting of efficiency spread gap and technological gap, but we used a Simar and Zelenyuk (2006) test instead of a Kruskal-Wallis test as an improvement on testing procedure. The results of this test are presented in Tables 5 and 6.[11] The findings of the tests show that there are significant differences in inefficiencies between all the regions studied.

*Table 5.* *Results of Simar and Zelenyuk (2006) adapted Li test (test statistic and p-values) to compare the efficiency spreads within groups*

|  | Asia-Pacific | | Europe | | USA-Canada | |
|---|---|---|---|---|---|---|
|  | Statistic | p-value | Statistic | p-value | Statistic | p-value |
| Asia-Pacific | - | - | 59.9369 | $\approx 0$ | 67.787 | $\approx 0$ |
| Europe |  |  | - | - | 5.4978 | $\approx 0$ |
| USA-Canada |  |  |  |  | - | - |

---

[11] The tests reported here are those run for all years together, due to space limitations. The results of the tests for each year can be obtained upon request.

*Table 6. Results of Simar and Zelenyuk (2006) adapted Li test (test statistic and p-values) to compare the relative position of the frontiers*

|  | Statistic | p-value |
|---|---|---|
| A - Asia-Pacific, B – Europe, C- USA-Canada | | |
| $CI_{NC}^{A}(Y^{A})$ vs $CI_{NC}^{R}(Y^{A})$ | 94.6612 | $\approx 0$ |
| $CI_{NC}^{R}(Y^{B})$ vs $CI_{NC}^{B}(Y^{B})$ | 3.5664 | 0.001 |
| $CI_{NC}^{R}(Y^{C})$ vs $CI_{NC}^{C}(Y^{C})$ | 11.4995 | $\approx 0$ |

## 5. Conclusions

The paper analyzed the performance of food and beverage manufacturing firms with regard to their CSR practices over the period 2009–2018. To differentiate from previous research, we analyzed CSR engagement of food and beverage firms worldwide, represented by the regions of Europe, the United States and Canada, and the Asia-Pacific. We applied the method of a DEA-based composite indicator (that is BoD model) in order to aggregate three dimensions of CSR: environmental, social, and governance. We further modified an original BoD model to account for non-convexities through an FDH approach and using this enhanced indicator we developed the indices to compare CSR performance of firms between regions and over time through the application of a Camanho-Dyson index. The study found differences in CSR efficiency between food and beverage firms in the regions considered, with USA-Canadian firms tending to perform best, followed by European firms, and Asian-Pacific firms achieving the worst efficiency results. The study also found some catching up process in the performance taking place in some years of the analysis, with Asia-Pacific catching up on Europe and USA-Canada, and Europe on USA-Canada.

Measuring CSR performance is necessary to guide sustainability improvements. The results of this study could be of interest to food and beverage firms' managers, CSR analysts, and policy makers in terms of how firms perform with regard to CSR, how this performance changes over time, and how it compares between firms in different regions. The results provide firms with a better understanding of CSR performance, which could encourage firms to develop their business operations towards more responsible and sustainable practices. Extensions of the results of this study could include the applications of ranking methods of firms' CSR efficiency or extensions in the BoD model in order to account for the fact that CSR indicators are imprecise in nature. Also, the development of composite indicators that would measure changes over time (productivity, efficiency, and technology) within FDH approach is a potential line of future research.

**ACKNOWLEDGMENTS**

Early version of this paper was presented at 15th International European Forum (Igls-Forum) and EAAE Seminar (Proceedings in System Dynamics and Innovation in Food Networks 2021). Financial support for this article from the National Science Centre in Poland (decision number DEC-2016/23/B/HS4/03398) is gratefully acknowledged. The calculations of adapted Li test were made at the Wroclaw Centre for Networking and Supercomputing (www.wcss.wroc.pl), grant no. 286. L. Ortiz thanks the grant PID2019-105952GB-I00 funded by Ministerio de Ciencia e Innovación/ Agencia Estatal de Investigación /10.13039/501100011033.